\documentclass[conference]{IEEEtran}
\usepackage{setspace}
\usepackage{amssymb}
\usepackage{amsmath}
\usepackage{bbm}
\usepackage{amsthm}
\usepackage{mathtools}
\usepackage{stmaryrd}
\usepackage{algorithmic}
\usepackage{tikz}
\usetikzlibrary{shapes}
\usepackage{tabularx}
\usepackage{multirow}
\usepackage{arydshln}
\usepackage{lscape}  
\usepackage{enumitem}
\usepackage[justification=raggedright]{caption}	
\usepackage{acro}
\acsetup{first-style=short}
\usepackage{array}
\newcolumntype{x}[1]{>{\centering\arraybackslash\hspace{0pt}}p{#1}}
\usepackage{color}
\usepackage[autostyle, english = american]{csquotes}
\MakeOuterQuote{"}

\newcounter{Q}
\newcounter{TD}

\DeclareMathOperator\md{\hspace{-8pt}\mod\hspace{-3pt}\Delta}
\DeclareMathOperator\itv{\left[-\tfrac{\Delta}{2},\tfrac{\Delta}{2}\right)}
\DeclareMathOperator{\sgn}{sgn}
\newcommand\given[1][]{\:#1\vert\:}

\newtheorem{thm}{Theorem}[section]
\newtheorem{lem}{Lemma}

\theoremstyle{definition}

\ifCLASSINFOpdf
  \graphicspath{{Figures/}}
  \DeclareGraphicsExtensions{.pdf,.jpeg,.png}
\else

\fi

\usepackage{subfigure}  
\usepackage{epstopdf}
\usepackage{caption}

\usepackage{amsmath}
\DeclareMathOperator*{\argmin}{argmin\,}
\DeclareMathOperator*{\argmax}{argmax\,}

\begin{document}

\title{Uncoded Binary Signaling through Modulo AWGN Channel}

\author{Gizem~Tabak and~Andrew~Singer\\
Coordinated Science Laboratory, Department of Electrical and Computer Engineering\\
University of Illinois at Urbana–Champaign, Urbana, IL 61801 USA}

\maketitle

\begin{abstract}
Modulo-wrapping receivers have attracted interest in several areas of digital communications, including precoding and lattice coding. The asymptotic capacity and error performance of the modulo AWGN channel have been well established. However, due to underlying assumptions of the asymptotic analyses, these findings might not always be realistic in physical world applications, which are often dimension- or delay-limited. In this work, the optimum ways to achieve minimum probability of error for binary signaling through a scalar modulo AWGN channel is examined under different scenarios where the receiver has access to full or partial information. In case of partial information at the receiver, an iterative estimation rule is proposed to reduce the error rate, and the performance of different estimators are demonstrated in simulated experiments.
\end{abstract}

\maketitle
\let\thefootnote\relax\footnotetext{Appeared at the 54th IEEE Asilomar Conference on Signals, Systems \& Computers. \vspace{-.4cm}}
\section{Introduction}
Modulo wrapping is encountered in various fields of digital communications. In precoding literature, modulo operation is used to cancel out known intersymbol interference at the transmitter \cite{tomlinson1971new}. In information theory literature, nested lattice codes \cite{Erez2004} and lattice Gaussian codes \cite{Ling2014} are proven to achieve additive white Gaussian noise (AWGN) channel capacity asymptotically under certain conditions. Both works, and many more in lattice coding literature, utilize modulo operation on lattices to encode, quantize or decode data. Recently, modulo wrapping receiver front-ends are used in analog-to-digital conversion to decrease distortion or to reduce number of quantization bits \cite{Ordentlich2018}. 

The asymptotic capacity and the error performance of the modulo channel have been well studied under the existing coding literature \cite{Erez2004, DavidForney2000a, Zamir2009}. The theoretical results, which are as impressive as achieving AWGN channel capacity asymptotically, are built on assumptions that are valid in high dimensions or long data block lengths. These assumptions might not always find correspondance in real physical applications which are often delay-limited and mostly built with hardware that operates in low dimensions. Moreover, in \cite{Ordentlich2018}, it is demonstrated that already-existing voltage-controlled oscillator (VCO) ADCs can be utilized as modulo receiver front-ends. Hence, the realization of modulo channels in physical, practical applications is feasible. Therefore, the non-asymptotical communication performance of modulo channels in low dimensions is of increased interest. In this work, the aim is to develop an optimum way (in terms of decoding error) for binary signaling through a scalar $\md$ AWGN channel with a power constraint, analyze the performance with different decision rules based on available information at the receiver, and propose an estimator to improve performance when the information at the receiver is lacking. 

In Section \ref{sec:2channel}, the transmission and decoding scheme over $\md$ AWGN channel will be introduced. In Section \ref{sec:3withprior}, the optimum transmission scheme and the performance of the communication will be derived when the prior information is available at the receiver. In Section \ref{sec:4noprior}, first, the optimum transmission scheme and the decoding performance with the blind decision rule will be derived when the prior information is not available at the receiver. Then, an iterative estimation rule will be proposed to improve the performance. The error rates of different estimators will be demonstrated on simulated experiments in Section \ref{sec:5sims}, and the findings will be concluded in \ref{sec:6conc}.
\vspace{-.15cm}
\section{Memoryless Mod-$\Delta$ Channel and Decoding}\label{sec:2channel}
\vspace{-.1cm}
In this paper, the following communication system will be considered: $N$ uncoded, independent and identically distributed (i.i.d.) sequence of data bits $\textbf{a}_0^{N-1} = (a_0,\dots,a_n,\dots,a_{N-1})\in \mathbb{Z}_2^N$ are mapped into symbols $\textbf{x}_0^{M-1} = (x_0,\dots,x_n,\dots,x_{M-1})\in \mathcal{S}^M$ to be transmitted through a modulo AWGN channel. The $\md$ AWGN channel is defined with the input-output relation
\begin{equation*}
    \tilde{y}_n = y_n \md = (x_n + w_n) \md
\end{equation*}
where $w_n$ are i.i.d. $\sim f_W(w) = \mathcal{N}(0, \sigma^2)$ noise samples. The modulo operation $\mod \Delta : \mathbb{R}\rightarrow \left[-\frac{\Delta}{2}, \frac{\Delta}{2}\right)$ is defined as
\begin{equation*}
    x \md = a \Leftrightarrow x = a+k\Delta, k \in \mathbb{Z} \text{ and } a \in \left[-\tfrac{\Delta}{2}, \tfrac{\Delta}{2}\right).
\end{equation*}
In the binary signaling problem, the symbols are given by
\begin{equation}\label{eq:2-symbols}
    x_{n} = 
    \begin{cases}
    h_0 \text{,   } a_n = 0\\
    h_1 \text{,   } a_n = 1
    \end{cases}
\end{equation}
where $\mathcal{S} = \{h_0, h_1\}$ and the power is assumed $P\geq {\Delta^2}/{2}$. 

The objective is to find the optimum mapping in \eqref{eq:2-symbols} that minimizes the probability of error with minimum power $P$ when recovering the sequence of data bits $\textbf{a}_0^{N-1}$ given the modulo-received samples $\tilde{\textbf{y}}_0^{N-1}$. More formally, minimizing the sequence error rate (SER) can be formulated as
\begin{align}\label{eq:2-MAP1}
    \hat{\textbf{a}}_0^{N-1} 
    = & \argmax_{\textbf{a}\in\mathbb{Z}_2^N} P(\textbf{a}_0^{N-1} = \textbf{a} | \tilde{\textbf{y}}_0^{N-1})\nonumber\\
    = & \argmax_{\textbf{a}\in\mathbb{Z}_2^N} P(\tilde{\textbf{y}} | \textbf{x} = \textbf{h}_\textbf{a}) P(\textbf{x} = \textbf{h}_\textbf{a})
\end{align}
where the $n$-th element of $\textbf{h}_\textbf{a}\in \mathbb{R}^N$ is $h_{a_n}$. The subscripts and superscripts are omitted for the sake of simplicity. 

If the priors $P(\textbf{x} = \textbf{h}_\textbf{a})$ are known at the receiver, then \eqref{eq:2-MAP1} corresponds to the maximum \textit{a posteriori} (MAP) estimation. If the priors are not known, then the maximum likelihood (ML) estimation 
\begin{align*}
    \hat{\textbf{a}}_0^{N-1} = \argmax_{\textbf{a}\in\mathbb{Z}_2^N} P(\tilde{\textbf{y}} | \textbf{x} = \textbf{h}_\textbf{a})
\end{align*}
is used.
\vspace{-.2cm}
\section{Modulation and Decoding with Prior Information}\label{sec:3withprior}
If the prior information is available at the receiver, than the MAP estimator in \eqref{eq:2-MAP1} would be used for decoding to minimize the probability of error on the $\md$ AWGN channel. Knowing the decision rule, the transmitter can optimize the symbols to be sent through the channel to minimize the probability of error and also the transmit power.

Define $f_W^{\Delta}\colon \mathbb{R} \to \mathbb{R^{+}}$, such that $x \mapsto f_W^{\Delta}(x) =  \sum_{k=-\infty}^\infty f_W(x + k\Delta)$ where $W \sim \mathcal{N}(0, \sigma^2)$. Then, the MAP estimation \eqref{eq:2-MAP1} yields
\begin{align*}
    \hat{a}_n   
    & = \argmax_{{a}\in \mathbb{Z}_2} P(\tilde{y}_n|a_n=a)P(a_n=a)\nonumber\\
    & = \argmax_{{a}\in \mathbb{Z}_2} \!\sum_{k=-\infty}^\infty f_{W}(\tilde{y}_n - h_a + k\Delta)  P(a_n=a)\nonumber\\
    & = \argmax_{{a}\in \mathbb{Z}_2} f_{W}^\Delta(\tilde{y}_n \!-\!h_a) P(a_n=a)
\end{align*}
for $n=0,\dots,N-1$, and the corresponding MAP decision rule, denoted as $\delta_{\pi_0}(\tilde{y}_n)$, becomes
\begin{align}\label{eq:2-NULRdec}
\delta_{\pi_0}(\tilde{y}_n ) 
    & \!=\! \begin{cases}
        0,  \pi_0 f_{W}^\Delta(\tilde{y}_n \!-\!h_0)\! \geq\! (1\!-\!\pi_0)f_{W}^\Delta(\tilde{y}_n \!-\!h_1)\\
        1,  \pi_0 f_{W}^\Delta(\tilde{y}_n \!-\!h_0)\!   < \! (1\!-\!\pi_0)f_{W}^\Delta(\tilde{y}_n \!-\!h_1)
        \end{cases}
\end{align}
where $\pi_0 = P(a_n=0)$.

The likelihood ratio for i.i.d. random bits ${a_n}$ is defined as
\begin{equation}\label{eq:2-NULR}
    L_{\pi_0}(a_n|\tilde{y}_n) 
    = \dfrac{P(a_n=0|\tilde{y}_n)}{P(a_n=1|\tilde{y}_n)}  
    = \dfrac{P(\tilde{y}_n| a_n=0)}{P(\tilde{y}_n|a_n=1)}\dfrac{\pi_0}{1-\pi_0} 
\end{equation}
and the MAP decision rule in \eqref{eq:2-NULRdec} can be written as
\begin{align}\label{eq:2-map-test}
    \hat{a}_n = \delta_{\pi_0} (\tilde{y}_n) 
    & = 
        \begin{cases}
        0, \:\: L_{\pi_0}(a_n|\tilde{y}_n) \geq 1\\
        1, \:\: L_{\pi_0}(a_n|\tilde{y}_n) < 1
        \end{cases}
\end{align}
and the probability of bit error is
\begin{align*}
    P_{e}(\delta_{\pi_0})
    = & P(\hat{{a}}_n = 1 | a_n = 0)P(a_n = 0)\\
      & + P(\hat{{a}}_n = 0 | a_n = 1)P(a_n = 1)
\end{align*}
\subsection{Uniform Priors}
\begin{figure}[t]
    \centering\hspace{-.5cm}
    \includegraphics[width=.47\textwidth]{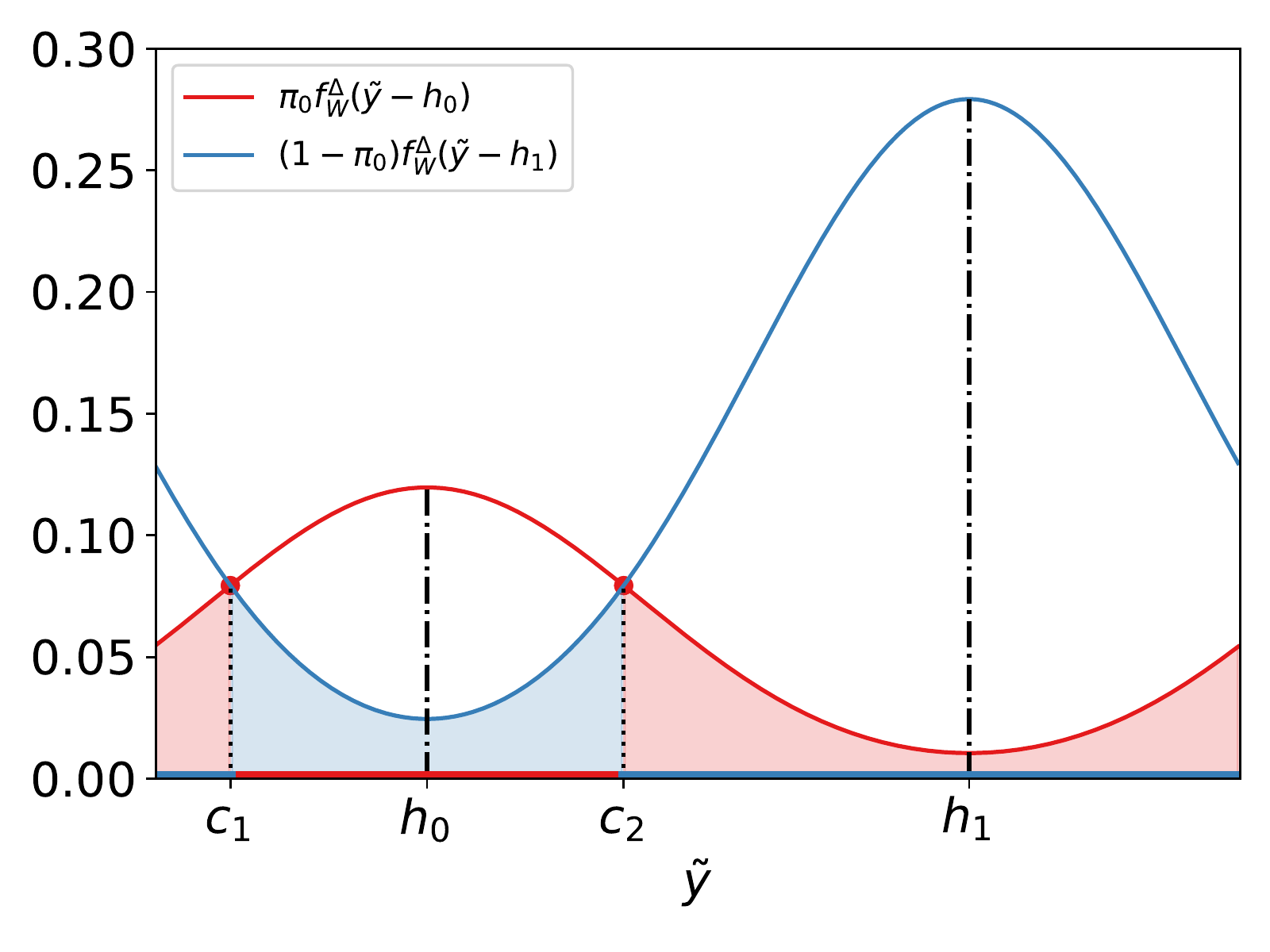}
    \vspace{-.3cm}
    \caption{Likelihood functions and MAP decision regions. Probability of error with MAP decision rule, $P_e(\delta_{\pi_0})$, corresponds to the shaded area under the curves}
    \label{fig:MAPregions}
    \vspace{-.3cm}
\end{figure}

Now, assume that the data bits are not only i.i.d., but also uniformly distributed (UID). In that case, the likelihood ratio in \eqref{eq:2-NULR} becomes
\begin{equation*}
    L(a_n|\tilde{y}_n) 
    = \dfrac{P(\tilde{y}_n| a_n=0)}{P(\tilde{y}_n|a_n=1)}
\end{equation*}
and it results in the decision rule $\delta(\tilde{y}_n)$
\begin{equation*}
    \hat{a}_n = \delta (\tilde{y}_n) = 
        \begin{cases}
        0, \:\: L(a_n|\tilde{y}_n) \geq 1\\
        1, \:\: L(a_n|\tilde{y}_n) < 1
        \end{cases}
\end{equation*}

Note that $f_{W}^\Delta(\tilde{y}_n - h_a) = f_{W}^\Delta(\tilde{y}_n - h_a + k\Delta)\: \forall k\in \mathbb{Z}$, and 
$\{h_{a} = h + k\Delta | \: k \in \mathbb{Z}\}$ yield the same decision and hence the same probability of error. Therefore, the transmitter can restrict the symbols to the interval $\itv$ to reduce the average transmit power $P$. Furthermore, the transmitter can set the values for $h_{a_n}$ appropriately to minimize the probability of error, $1 - P(\hat{\textbf{a}}=\textbf{a}) = 1 - \prod P(\hat{a}_n=a_n)$, for given $\Delta$ and $\sigma$. This claim is formalized in Theorem \ref{thm:2.1} and the proof is given in Appendix \ref{proof:thm1}.

\begin{lem}\label{lem:2-1inters}
If $z\in\mathbb{R}$ and $\mathcal{H}=\left\{c \in \itv \middle| f_W^{\Delta}(c) = z\right\}$, then $\left|\mathcal{H}\right| \leq 2$.
\end{lem}

\begin{lem}\label{lem:2-2sol}
Let $a$ be UID over $\mathbb{Z}_2$, $h_0 \neq h_1$, and define $\mathcal{C}=\{c \in \itv \given[\big] L(a=0|c) = L(a=1|c)\}$. Then, $\mathcal{C} = \left\{ \tfrac{h_0 + h_1}{2}, \tfrac{h_0 + h_1}{2} +  \sgn\left({\frac{h_0+h_1}{2}}\right)\tfrac{\Delta}{2} \right\}$ where $\sgn$ denotes the modified sign function so that $\sgn (0) = -1$.
\end{lem}

\begin{thm}\label{thm:2.1}
Let $a_n$ be UID over $\mathbb{Z}_2$ and $x_n$ the symbols representing $a_n$ as defined in \eqref{eq:2-symbols}. Then, $h_0 = -\tfrac{\Delta}{4}$ and $h_1 = \tfrac{\Delta}{4}$ yields the minimum probability of error, $P_e^*$, while minimizing the average power $P = \mathbb{E}[x_n^2]$ for $P_e^*$.
\end{thm}

If the symbol mapping in Theorem \ref{thm:2.1} is used at the transmitter, then the probability of bit error becomes
\begin{align*}
    P_{e} 
    & = \Phi\left(\frac{\Delta}{4\sigma}\right) - \Phi\left(\frac{3\Delta}{4\sigma}\right)
\end{align*}
where $\Phi\left(\frac{x}{\sigma}\right) =  \sum_{k=-\infty}^\infty Q\left(\frac{x + k\Delta}{\sigma}\right)$. Note that the probability of error with optimum transmit symbols depends on the ratio $\Delta/\sigma$.
\subsection{Nonuniform Priors}
If the data bits are not UID but the prior $\pi_0$ is known at the receiver, then the decision rule in \eqref{eq:2-map-test} will be used to decode the received symbols. In Fig. \ref{fig:MAPregions}, the likelihood functions, decision regions (horizontal bars) and the probability of error (shaded area) is demonstrated for $\Delta/\sigma=5$. To derive the symbol mapping that will yield the minimum probability of error with minimum average power when the priors are available at the receiver, first, define
\begin{align}
    P_1 
    & = \pi_0\left(\Phi(c_1-h_0) - \Phi(c_2-h_0)\right) \label{eq:P1-1stline}\\
    & + (1-\pi_0)\left(\Phi(c_2-h_1) - \Phi(c_1-h_1+\Delta)\right) \label{eq:P1-2ndline}\\
    P_2 
    & = \pi_0\left(\Phi(c_2-h_0) - \Phi(c_1-h_0+\Delta)\right) \nonumber\\
    & + (1-\pi_0)\left(\Phi(c_1-h_1) - \Phi(c_2-h_1)\right)\nonumber
\end{align}
where $\mathcal{C} = \{c_1, c_2\}$ is defined as in Lemma \ref{lem:2-1inters}. Note that \eqref{eq:P1-1stline} corresponds to the red shaded region and \eqref{eq:P1-2ndline} corresponds to the blue shaded region in Fig. \ref{fig:MAPregions}. The, the probability of error can be written as
\begin{equation*}
    P_{e}(\delta_{\pi_0})\! =\! 
    \begin{cases}
    P_1, & \!\!\pi_0f_W^{\Delta}(-\frac{\Delta}{2}-h_0)>(1-\pi_0)f_W^{\Delta}(-\frac{\Delta}{2}-h_1) \\
    P_2, & \!\!\pi_0f_W^{\Delta}(-\frac{\Delta}{2}-h_0)\leq(1-\pi_0)f_W^{\Delta}(-\frac{\Delta}{2}-h_1)
    \end{cases}
\end{equation*}
\begin{thm}\label{thm:2.2}
Let $a_n$ be i.i.d. over $\mathbb{Z}_2$ and $x_n$ the symbols representing $a_n$ as defined in \eqref{eq:2-symbols}. Then, $h_0 = -\tfrac{\Delta}{4}$ and $h_1 = \tfrac{\Delta}{4}$ yields the minimum probability of error with decision rule \eqref{eq:2-NULRdec} while minimizing the average power $P = \mathbb{E}[x_n^2]$.
\end{thm}
\begin{figure}[t]
    \centering
    \hspace{-.5cm}
    \includegraphics[width=.47\textwidth]{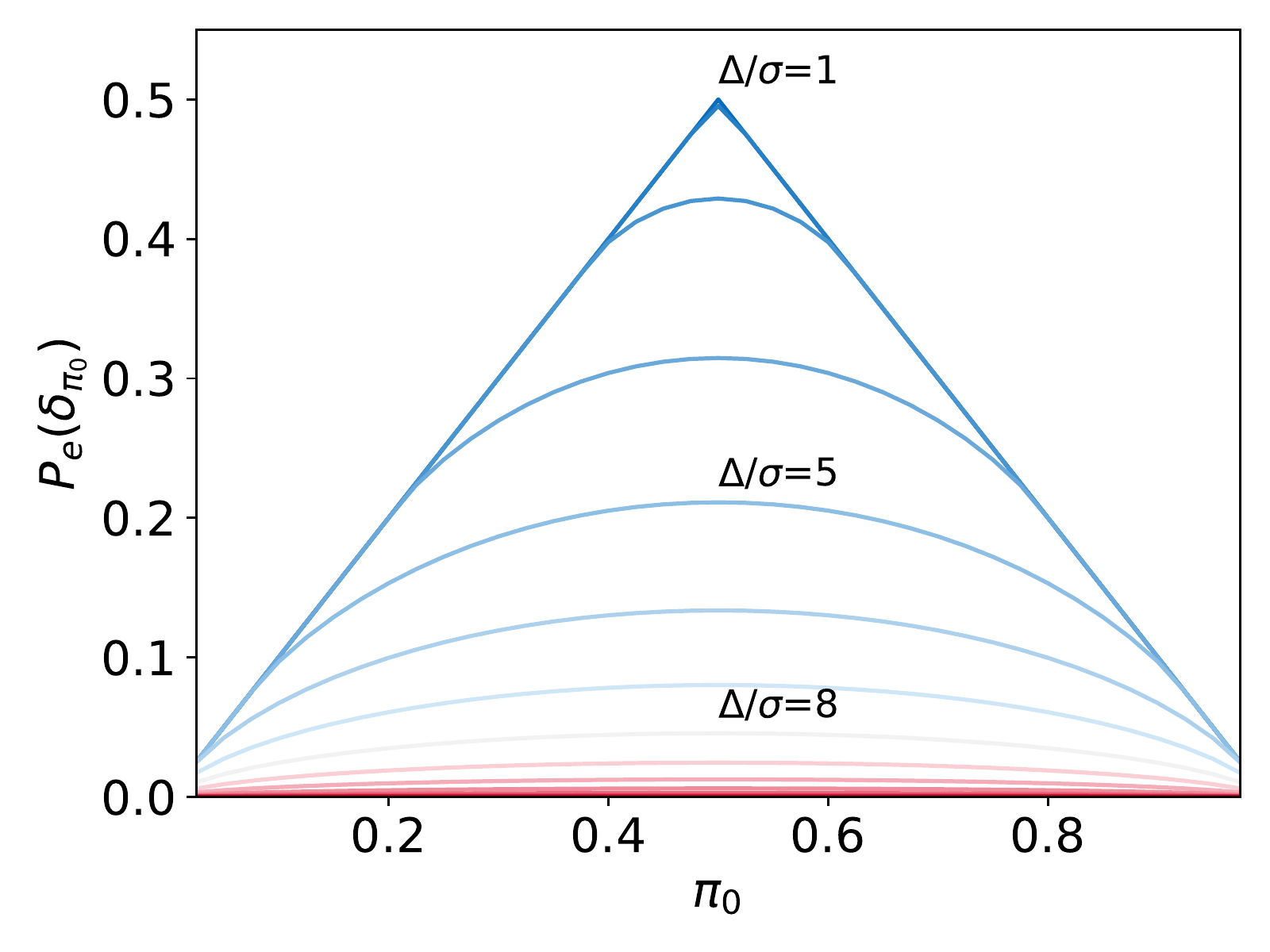}
    \vspace{-.3cm}
    \caption{Probability of error with MAP decision rule $P_e(\delta_{\pi_0})$ with different priors $\pi_0$ for different $\Delta/\sigma$ ratios} 
    \label{fig:map_pe_vs_p0}
    \vspace{-.4cm}
\end{figure}
Then, the corresponding probability of error becomes
\begin{align}\label{eq:MAP_pe}
    & P_e(\delta_{\pi_0}) = \\
    & \begin{cases}
    \hfil 1-\pi_0
    \!\!\!\!\!\!\!\!&\!\!\!\!\!, \pi_0 > \frac{1}{1+\tau}\\
    (1-\pi_0)\left(\Phi\left(\frac{l_{\pi_0}}{\sigma}\right) - \Phi\left(\frac{\Delta-l_{\pi_0}}{\sigma}\right)\right) \\
    + \pi_0\left(\Phi\left(\tfrac{\Delta/2-l_{\pi_0}}{\sigma}\right) - \Phi\left(\tfrac{\Delta/2+l_{\pi_0}}{\sigma}\right)\right)
    \!\!\!\!\!\!\!\!&\!\!\!\!\!\hspace{-.1cm}, \frac{1}{1+\tau} \geq \pi_0 > 0.5\\
    (1-\pi_0)\left(\Phi\left(\tfrac{\Delta/2-l_{\pi_0}}{\sigma}\right) - \Phi\left(\tfrac{\Delta/2+l_{\pi_0}}{\sigma}\right)\right) \\
    +  \pi_0\left(\Phi\left(\tfrac{l_{\pi_0}}{\sigma}\right) - \Phi\left(\tfrac{\Delta-l_{\pi_0}}{\sigma}\right)\right)
    \!\!\!\!\!\!\!\!&\!\!\!\!\!\hspace{-.1cm}, \frac{\tau}{1+\tau} \leq \pi_0 < 0.5\\
    \hfil \pi_0 
    \!\!\!\!\!\!\!\!&\!\!\!\!\!\hspace{-.1cm}, \pi_0 < \frac{\tau}{1+\tau}
    \end{cases}\nonumber
\end{align}
where $\tau = \tfrac{f_W^{\Delta}(\Delta/2)}{f_W^{\Delta}(0)}$ and $\mathcal{C} = \{c_1, c_2\} = \{h_0-l_{\pi_0}, h_0+l_{\pi_0}\}$. Note that the probability of error with the MAP decision rule depends on $\Delta/\sigma$ and also on $\pi_0$ (Fig. \ref{fig:map_pe_vs_p0}). For smaller values of $\Delta/\sigma$, the rule yields higher probability of error, which is caused by the information that is lost by the modulo wrapping operation.

\section{Modulation and Decoding without Prior Information}\label{sec:4noprior}
\subsection{ML Decision Rule}
If the prior $\pi_0$ is not available at the receiver, then the ML decision rule $\delta$ is 
\begin{align}\label{eq:up-ml-rule}
    \delta(\tilde{y}_n) 
    & = \begin{cases}
        0, \:\: f_{W}^\Delta(\tilde{y}_n -h_0) \geq f_{W}^\Delta(\tilde{y}_n -h_1)\\
        1, \:\: f_{W}^\Delta(\tilde{y}_n-h_0) < f_{W}^\Delta(\tilde{y}_n-h_1)
        \end{cases}
\end{align}
and the resulting probability of error is
\begin{align}\label{eq:up-ml-pe}
    P_e(\delta) \!
    & = \!\begin{cases}
    \Phi\left(\frac{h_1-h_0}{2\sigma}\right) - \Phi\left(\frac{h_1-h_0}{2\sigma}+\frac{\Delta}{2\sigma}\right), & \frac{h_0+h_1}{2\sigma}\geq0\\
    \Phi\left(\frac{h_1-h_0}{2\sigma}-\frac{\Delta}{2\sigma}\right) - \Phi\left(\frac{h_1-h_0}{2\sigma}\right), & \frac{h_0+h_1}{2\sigma}<0
    \end{cases}
\end{align}

\begin{figure}[t]
    \centering\hspace{-.3cm}
    \includegraphics[width=.47\textwidth]{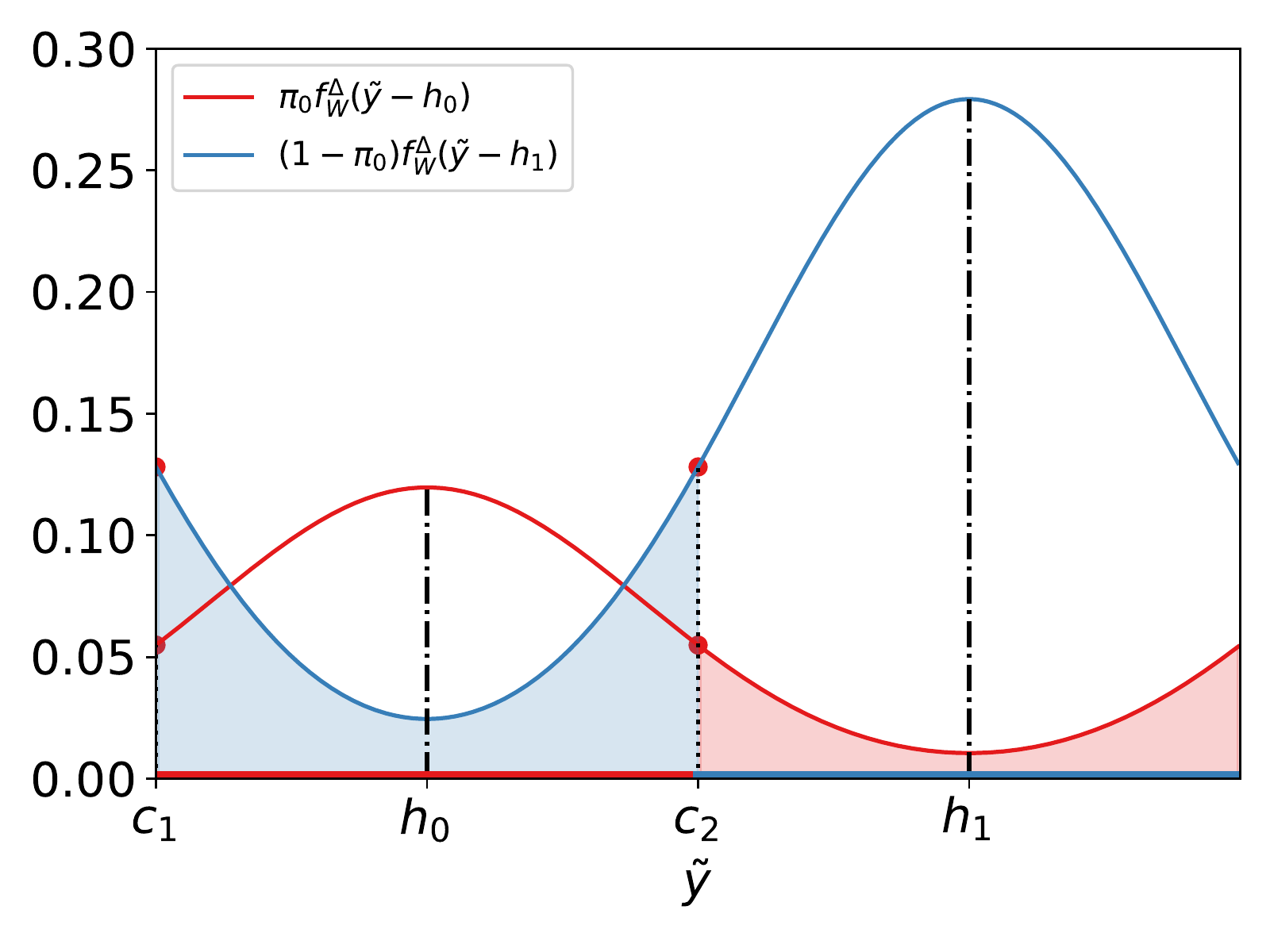}
    \vspace{-.3cm}
    \caption{Likelihood functions and ML decision regions. Probability of error with ML decision rule, $P_e(\delta)$, corresponds to the shaded area under the curves}
    \label{fig:MLregions}
    \vspace{-.4cm}
\end{figure}

It is evident from Theorem \ref{thm:2.1} that $h_0=-\tfrac{\Delta}{4}$ and $h_1=\tfrac{\Delta}{4}$ minimizes \eqref{eq:up-ml-pe}, yielding
\begin{equation}\label{eq:up-ml-pefinal}
    P_e(\delta) = \Phi\left(\frac{\Delta}{4\sigma}\right) - \Phi\left(\frac{3\Delta}{4\sigma}\right)
\end{equation}
The decision regions of ML rule  are demonstrated in Fig. \ref{fig:MLregions} on the same setup as in Fig. \ref{fig:MAPregions}. Note that $P_e(\delta_{\pi_0}) \leq P_e(\delta)$, and the equality holds when $\pi_0 = 0.5$. 
\subsection{Prior Estimation}
As it is also evident from \eqref{eq:up-ml-pefinal} and from the shaded regions in Fig. \ref{fig:MLregions}, blind ML decision rule results in higher probability of error compared to the MAP decision rule. One possible way to decrease the error rate when priors are unknown at the receiver is to use a two-step decision process: First, apply the ML rule and estimate the prior from the decisions. Then, use the MAP decision rule with the estimated prior. 

First, note that the decision rule $\delta(\tilde{y}_i)$ yields to a Bernoulli random variable with
\begin{align*}
    P(\delta(\tilde{y}_i) = 1) 
    = & P(\delta(\tilde{y}_i) = 1 | a_i = 0)P(a_i=0) \nonumber\\
      & + P(\delta(\tilde{y}_i) = 1 | a_i = 1)P(a_i=1) \nonumber\\
    = & \pi_0 P_e(\delta) + (1-\pi_0)(1- P_e(\delta) \coloneqq p_b
\end{align*}
and the sum of ML decisions is a binomial random variable, denoted as $N_{\delta}=\sum_{i=1}^N\delta(\tilde{y}_i)\sim \text{Binom}(p_b)$. Note that
\begin{equation*}
    \mathbb{E}[N_{\delta}] = Np_b = N\left(\pi_0 P_e(\delta) + (1-\pi_0)(1- P_e(\delta)\right)
\end{equation*}
and the prior $\pi_0$ can be estimated empirically as
\begin{align*}
    \hat{\pi}_0 & = \dfrac{\tfrac{1}{N}\sum_{i=1}^N\delta(\tilde{y}_i) - (1-P_e(\delta))}{2P_e(\delta)-1},
\end{align*}
which results in the decision rule
\begin{align}\label{eq:2-est-dec}
\delta_{\hat{\pi}_0}(\tilde{y}_n ) 
    & \!=\! \begin{cases}
        0,  \hat{\pi}_0 f_{W}^\Delta(\tilde{y}_n \!-\!h_0)\! \geq\! (1\!-\!\hat{\pi}_0)f_{W}^\Delta(\tilde{y}_n \!-\!h_1)\\
        1,  \hat{\pi}_0 f_{W}^\Delta(\tilde{y}_n \!-\!h_0)\!   < \! (1\!-\!\hat{\pi}_0)f_{W}^\Delta(\tilde{y}_n \!-\!h_1)
        \end{cases}
\end{align}
and yields $P_e(\delta_{\hat{\pi}_0})$. 
%

\begin{figure}[t]
    \hspace{-.5cm}
    \includegraphics[width=.5\textwidth]{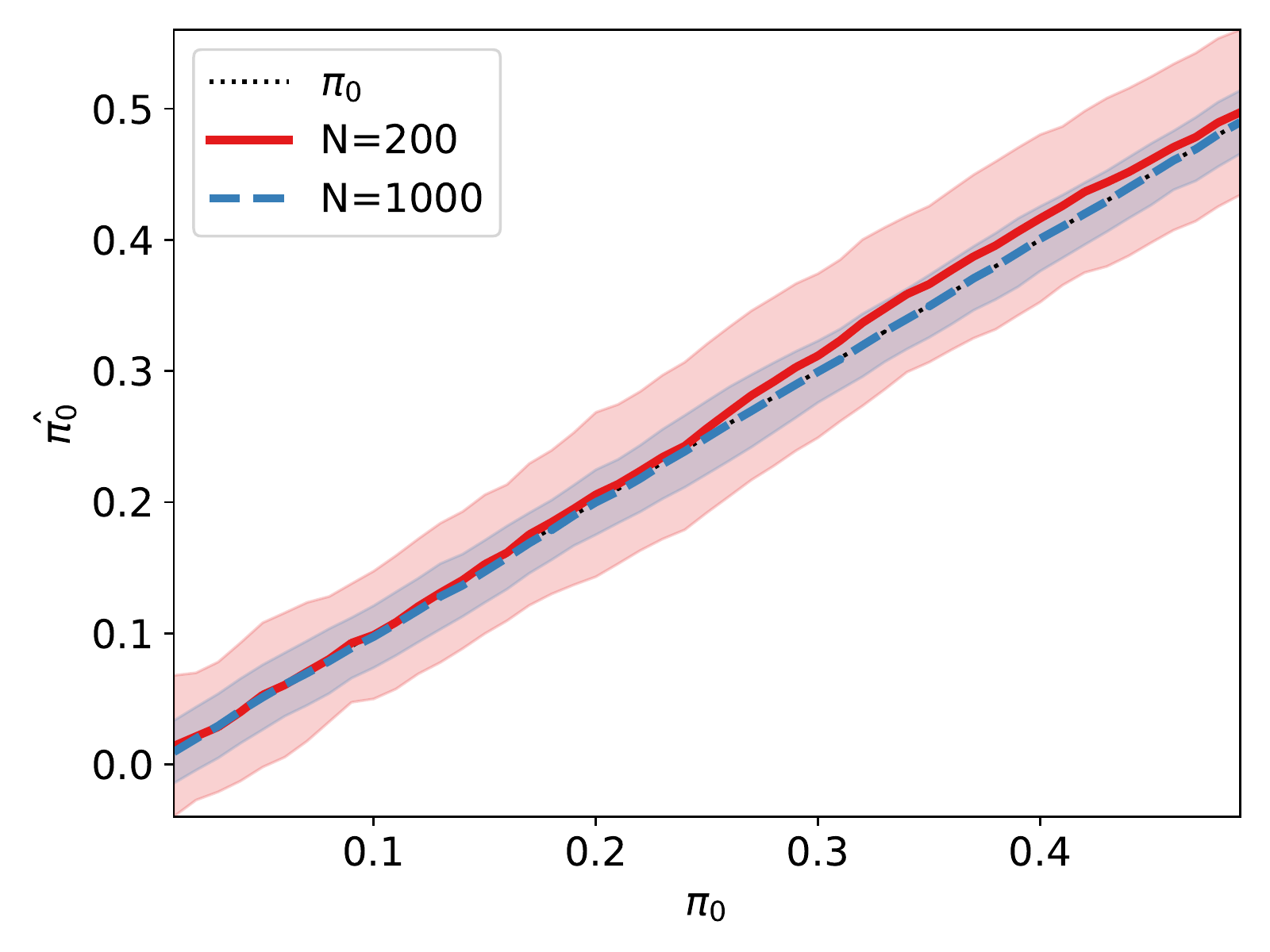}
    \caption{Real ($\pi_0$) and estimated ($\hat{\pi}_0$) priors using $N=\{200, 1000\}$ symbols ($\Delta/\sigma=5$)}
    \label{fig:exp_p0hat}
    \vspace{-.3cm}
\end{figure}
    
\section{Experiments}\label{sec:5sims}
A set of simulated experiment are performed in order to demonstrate the decoding performance of different decision rules. In each experiment, $N$ random data bits are generated from a Bernoulli distribution $\text{Ber}(\pi_0)$ and mapped to symbols according to \eqref{eq:2-symbols}. Then, AWGN samples with $\mathcal{N}(0,1)$ are added to each symbol. Because $\Delta/\sigma$ ratio determines the performance, $\sigma$ is fixed and modulo wrapping is performed with different $\Delta$ values. For decoding the noisy, modulo-wrapped symbols, MAP, ML and estimated decision rules in \eqref{eq:2-NULRdec}, \eqref{eq:up-ml-rule} and \eqref{eq:2-est-dec} are used, respectively. The experiments are repeated $M=50$ times, unless otherwise noted.

In Fig. \ref{fig:exp_p0hat}, the estimated priors from different number of symbols are demonstrated. The lines correspond to the average and the shaded regions correspond to the standard deviation of error rate for $M$ experiments.  The plots are limited to $\pi_0\in \left[0, 0.5\right]$ because of symmetry. 

\begin{figure}[t]
\hspace{-.5cm}
    \includegraphics[width=.5\textwidth]{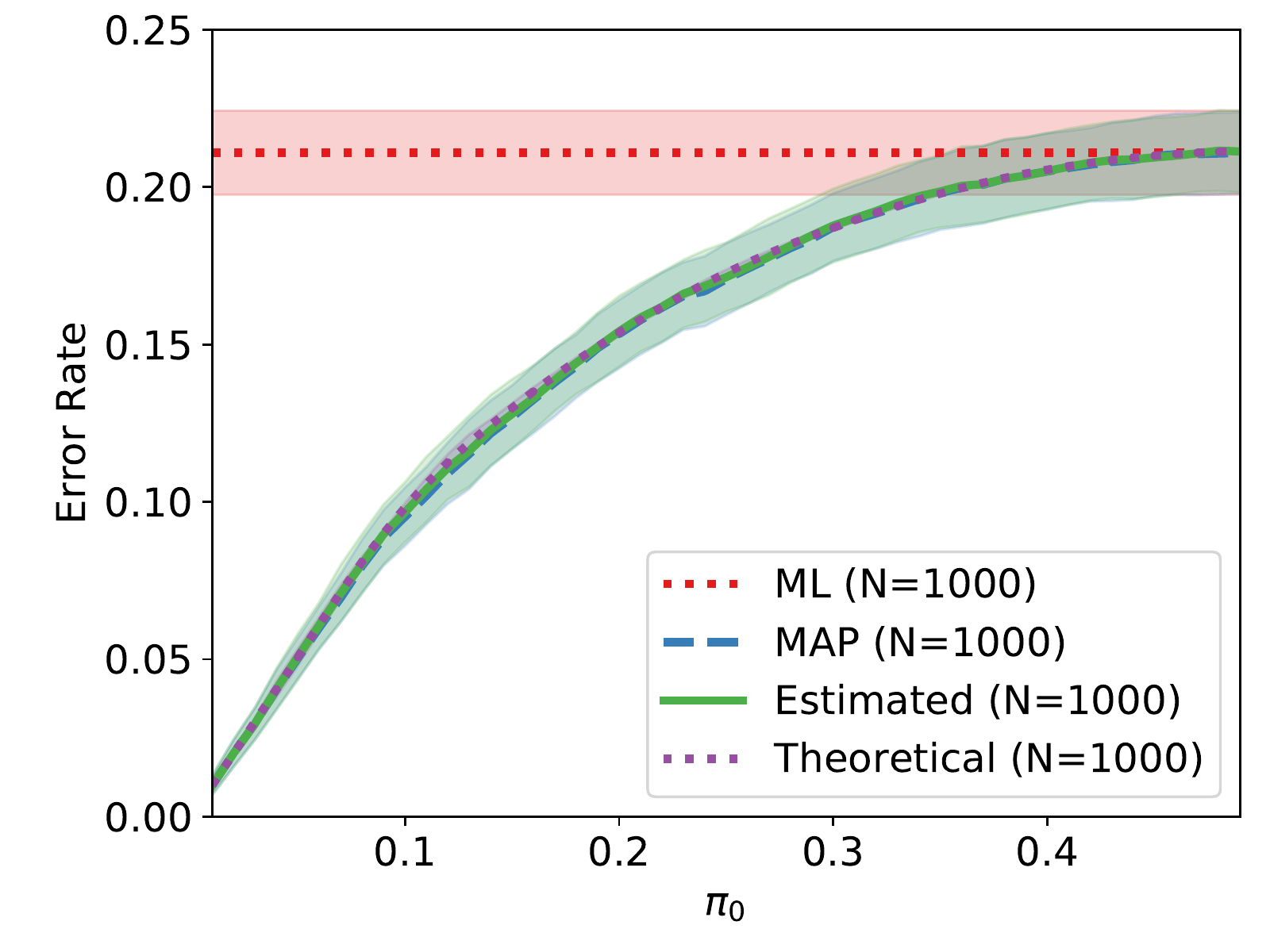}
    \caption{Empirical error rate with different decision rules ($\Delta/\sigma=5$)}
    \label{fig:exp_accuracy}
    \vspace{-.3cm}
\end{figure}

In Fig. \ref{fig:exp_accuracy}, the resulting error rate of three decision rules for $\Delta/\sigma=5$ are compared. The error rate of the estimated decision rule is close to MAP rule, which is in line with the theoretical probability of error in \eqref{eq:MAP_pe}. Both MAP rules with known and estimated priors yield lower error rates than ML rule when the bits are further from uniform, and they yield the same result as expected when $\pi_0=0.5$.

Although the MAP rule with estimated priors results in improved probability of error compared to ML rule for $\Delta/\sigma=5$, it is important to note that its performance inherently depends on the ML rule. If the decisions from ML rule is unreliable, e.g. $\Delta/\sigma$ is too small, then the prior estimate and hence the estimated MAP rule will not perform as well. However, the ratio $\Delta/\sigma$ being too small results in high decoding errors even when the priors are known, as it is seen in Fig. \ref{fig:map_pe_vs_p0}. Therefore, operating in low $\Delta/\sigma$ regime is not ideal to begin with.
\section{Conclusion}\label{sec:6conc}
Modulo receivers can provide several improvements in a communication problem. In this work, the optimum communication scheme through a $\md$ AWGN channel is explored under different scenarios for the available information at the receiver. It is demonstrated that the optimum binary signaling on a $\md$ AWGN channel to minimize the probability of decoding error is to use $\mathcal{S}=\{-\tfrac{\Delta}{2}, \tfrac{\Delta}{2}\}$ regardless of the information available at the receiver. This is in line with nested lattice codes in \cite{Zamir2009}, where $\Lambda=\Delta\mathbb{Z}$ and $\Lambda^\prime = \tfrac{\Delta}{2}\mathbb{Z}$.

Moreover, a modified, iterative decision rule is proposed instead of ML rule when the prior information is not available at the receiver. Although the modified decision rule might not achieve uniformly better performance than blind estimation, it improves performance when $\md$ channel is favorable.
\bibliographystyle{IEEEtran}
\bibliography{modcomms}

\begin{thebibliography}{1}
\providecommand{\url}[1]{#1}
\csname url@samestyle\endcsname
\providecommand{\newblock}{\relax}
\providecommand{\bibinfo}[2]{#2}
\providecommand{\BIBentrySTDinterwordspacing}{\spaceskip=0pt\relax}
\providecommand{\BIBentryALTinterwordstretchfactor}{4}
\providecommand{\BIBentryALTinterwordspacing}{\spaceskip=\fontdimen2\font plus
\BIBentryALTinterwordstretchfactor\fontdimen3\font minus
  \fontdimen4\font\relax}
\providecommand{\BIBforeignlanguage}[2]{{%
\expandafter\ifx\csname l@#1\endcsname\relax
\typeout{** WARNING: IEEEtran.bst: No hyphenation pattern has been}%
\typeout{** loaded for the language `#1'. Using the pattern for}%
\typeout{** the default language instead.}%
\else
\language=\csname l@#1\endcsname
\fi
#2}}
\providecommand{\BIBdecl}{\relax}
\BIBdecl

\bibitem{tomlinson1971new}
M.~Tomlinson, ``New automatic equaliser employing modulo arithmetic,''
  \emph{Electronics letters}, vol.~7, no.~5, pp. 138--139, 1971.

\bibitem{Erez2004}
U.~Erez, R.~Zamir, and S.~Member, ``{Achieving 0.5log(1+SNR) on the AWGN
  Channel With Lattice Encoding and Decoding},'' \emph{October}, vol.~50,
  no.~10, pp. 2293--2314, 2004.

\bibitem{Ling2014}
C.~Ling and J.~C. Belfiore, ``{Achieving AWGN channel capacity with lattice
  Gaussian coding},'' \emph{IEEE Transactions on Information Theory}, vol.~60,
  no.~10, pp. 5918--5929, 2014.

\bibitem{Ordentlich2018}
O.~Ordentlich, G.~Tabak, P.~K. Hanumolu, A.~C. Singer, and G.~W. Wornell, ``{A
  modulo-based architecture for analog-to-digital conversion},'' \emph{IEEE
  Journal on Selected Topics in Signal Processing}, 2018.

\bibitem{DavidForney2000a}
G.~{David Forney}, M.~D. Trott, and S.~Y. Chung, ``{Sphere-bound-achieving
  coset codes and multilevel coset codes},'' \emph{IEEE Transactions on
  Information Theory}, 2000.

\bibitem{Zamir2009}
R.~Zamir, ``{Lattices are everywhere},'' \emph{Information Theory and
  Applications Workshop, ITA 2009}, pp. 392--421, 2009.

\end{thebibliography}
\appendices
\section{Proof of Lemma \ref{lem:2-1inters}}\label{proof:lem1}
\begin{proof}
By definition, $z = f_W^{\Delta}(c) = \sum_{k=-\infty}^\infty f_W(c + k\Delta) = \sum_{k=-\infty}^\infty f_W(- c - k\Delta) = \sum_{k=\infty}^{-\infty} f_W(- c + k\Delta) = \sum_{k=-\infty}^{\infty} f_W(- c + k\Delta) = f^{\Delta}( - c)$, meaning $ - c \in \mathcal{H}$. If $0\notin \mathcal{H}$, there has to be even number of solutions and if there is odd number of solutions, then $0\in \mathcal{H}$. Now assume there are more than two solutions. Without loss of generality, one can assume $0 \leq c_0 < c_1 < \Delta/2$. Then, by Rolle's theorem, $\exists y \in [c_0,c_1] \colon \left(f^{\Delta}\right)^\prime(y) = 0$. Notice that the derivative $\left(f^{\Delta}\right)^\prime(y) = \sum_{k=-\infty}^{\infty} f^\prime_W(y + k\Delta)$ can be written as a sum over $m=\{0,\dots,\infty\}$ by pairing sum terms of $\{(k=m, k=-m-1)\}$.
Then, $\left(y +k   \Delta\right) e^{-\tfrac{1}{\sigma^2}(y +  k   \Delta)^2} 
    +\left(y-(k+1)\Delta\right) e^{-\tfrac{1}{\sigma^2}(y - (k+1)\Delta)^2} 
    <\left(2y-\Delta\right) e^{-\tfrac{1}{\sigma^2}(y + k\Delta)^2} < 0$, which means $\left(f^{\Delta}\right)^\prime(y) > 0$, which contradicts with Rolle's theorem. Then, there cannot be more than two solutions.
\end{proof}
\section{Proof of Lemma \ref{lem:2-2sol}}\label{proof:lem2}
\begin{proof}
If $a$ is uniform over $\mathbb{Z}_2$ then, $L(a=0|c) = L(a=1|c) \Rightarrow P(c|a=0) = P(c|a=1)$. Using Poisson's summation formula,
\begin{align*}
    P(c|a=0) 
    & = \dfrac{2}{\Delta} \sum_{k=-\infty}^\infty e^{-\tfrac{2\pi^2 \sigma^2 k^2}{\Delta^2}}\cos\left(2\pi k\frac{c - h_0}{\Delta}\right) \\
    P(c|a=1) 
    & = \dfrac{2}{\Delta} \sum_{k=-\infty}^\infty e^{-\tfrac{2\pi^2 \sigma^2 k^2}{\Delta^2}}\cos\left(2\pi k\frac{c - h_1}{\Delta}\right)
\end{align*}
which are equal when $c = \frac{n}{2k} \Delta + \frac{h_0+h_1}{2}, \: n, k \in \mathbb{Z}$. Note that $f_0^{\Delta}(x) = f^{\Delta}(x-h_0)$ is $\Delta-$periodic and symmetric around $h_0$. Then, $f_0^{\Delta}(\left( 2h_0 - c\right)\md) = f_0^{\Delta}(\left( c\right)\md)$ $=$ $f_0^{\Delta}(\left( c+h_0-h_1\right)\md)$ $=$ $f_0^{\Delta}(\left( h_0 + h_1 - c\right)\md) = \rho$. Lemma \ref{lem:2-1inters} entails two of the inputs of these four equations to be equal, which, combined with the constraint on $c$, yields $\tfrac{h_0 + h_1}{2}$ and $\tfrac{h_0 + h_1}{2} +  \sgn\left({\frac{h_0+h_1}{2}}\right)\tfrac{\Delta}{2}$.
\end{proof}
\section{Proof of Theorem \ref{thm:2.1}}\label{proof:thm1}
\begin{proof}
Since $a_n$ are independent, the probability of sequence error $P_e = 1 - \prod P(\hat{a}_n=a_n)$ is minimized when probability of bit error $P(\hat{a}_n\neq a_n)$ is minimized for all $n$. More formally,
\begin{equation*}
    \min_{h_0, h_1} P_e = 1 - \prod \left(1-\min_{h_0, h_1} P(\hat{a}_n \neq a_n) \right)
\end{equation*}
Without loss of generality, assume $h_0 < h_1$ and define $\mathcal{R}_0 = \{\tilde{y}\in \itv\given[\big] L(a|\tilde{y}) \geq 1 \}$ and $\mathcal{R}_1 = \{\tilde{y}\in \itv \given[\big] L(a|\tilde{y}) < 1 \}$. From Lemma \ref{lem:2-2sol}, it can be seen that 
\begin{itemize}
    \item If $\tfrac{h_0+h_1}{2}<0$, then
    \begin{align*}
        & c_0 = \tfrac{h_0+h_1}{2}, c_1 = \tfrac{h_0+h_1}{2} + \tfrac{\Delta}{2}\\
        & \mathcal{R}_1 = (c_0, c_1), \mathcal{R}_0 = \left[-\tfrac{\Delta}{2}, c_0\right] \cup \left[c_1, \tfrac{\Delta}{2}\right)
    \end{align*}
    \item If $\tfrac{h_0+h_1}{2}\geq 0$, then
    \begin{align*}
        & c_0 = \tfrac{h_0+h_1}{2} - \tfrac{\Delta}{2}, c_1 = \tfrac{h_0+h_1}{2}\\
        & \mathcal{R}_0 = (c_0, c_1), \mathcal{R}_1 = \left[-\tfrac{\Delta}{2}, c_0\right] \cup \left[c_1, \tfrac{\Delta}{2}\right)
    \end{align*}
\end{itemize}
Then, the probability of bit error becomes
\begin{align*}
    P_{e}
    = & P(\hat{{a}}_n = 1 | a_n = 0)P(a_n = 0) \nonumber\\
      & + P(\hat{{a}}_n = 0 | a_n = 1)P(a_n = 1)\nonumber\\
    = & \sum_{k=-\infty}^\infty \left(Q\left(\frac{\eta + k\Delta}{\sigma}\right) - Q\left(\frac{\eta+\frac{\Delta}{2} + k\Delta}{\sigma}\right)\right)
\end{align*}
where $\eta=\tfrac{h_1-h_0}{2}$. Setting $\frac{\partial P_{e}}{\partial \eta}=0$ yields
\begin{equation}\label{eq:2-deta}
    \sum_{k=-\infty}^\infty f_W\left(\eta+\frac{\Delta}{2} + k\Delta\right) = \sum_{k=-\infty}^\infty f_W\left(\eta + k\Delta\right)
\end{equation}{}
Lemma \ref{lem:2-2sol} dictates that the solutions to \eqref{eq:2-deta} are in the form of $2\eta +\tfrac{\Delta}{2} = k\tfrac{\Delta}{2}, k \in \mathbb{Z}$, and only $2\eta = h_1 - h_0 = \tfrac{\Delta}{2}$ satisfies the conditions for $h_0$ and $h_1$. For such $h_0$ and $h_1 = h_0 + \tfrac{\Delta}{2}$, the average power becomes $P = \mathbb{E}[x_n^2] = \tfrac{1}{2}h_0^2 + \tfrac{1}{2}(h_0+\tfrac{\Delta}{2})^2$. Then, $h_0^*$ and $h_1^*$ that minimize the average power is $h_0^* = -\frac{\Delta}{4} \text{ and } h_1^* = \frac{\Delta}{4}$.
\end{proof}
\section{Proof of Theorem \ref{thm:2.2}}\label{proof:thm2}
\begin{proof}
Without loss of generality, assume $h_1 = h_0 + \epsilon$, $0\leq\epsilon<\tfrac{\Delta}{2}-h_0$ and $c_1 \leq c_2$. Then, one can see that $\argmax_{\epsilon} P_{e} = 0$ and $\argmin_{\epsilon} P_{e} = \frac{\Delta}{2}$, by setting 
\begin{equation*}
    \dfrac{\partial P_{e}}{\partial \epsilon} = \pm \pi_0 \left(f_W^{\Delta}(c_2-h_0) - f_W^{\Delta}(c_1-h_0)\right) =0
\end{equation*}
and by incorporating the following definition of $c_1$ and $c_2$,
\begin{align*}
    f_W^{\Delta}(c_2-h_0) & = f_W^{\Delta}(c_1-h_0)\nonumber\\
    f_W^{\Delta}(c_2-h_1) & = f_W^{\Delta}(c_1-h_1)\nonumber
\end{align*}
into the derivative. Then, the $h_0$ and $h_1$ that minimize the power is $-\tfrac{\Delta}{4}$ and $\tfrac{\Delta}{4}$, as in Theorem \ref{thm:2.1}.
\end{proof}

\end{document}